\documentclass[twocolumn,aps,pra,longbibliography,superscriptaddress,nofootinbib,10pt]{revtex4-1}
\usepackage[latin1]{inputenc}
\usepackage{graphicx,dcolumn,bm}
\usepackage{subfigure}
\usepackage{multirow}
\usepackage{array}
\usepackage{arydshln}
\usepackage{color}
\usepackage[colorlinks,linkcolor=blue,citecolor=blue,hyperindex]{hyperref}
\usepackage{amsfonts}
\usepackage{amssymb}
\usepackage{amsmath}
\usepackage{latexsym}
\usepackage{simplewick}
\usepackage{epstopdf}
\usepackage{multirow}
\usepackage{float}
\usepackage[table]{xcolor}
\usepackage{multibib}
\usepackage{mathrsfs}
\usepackage[sans]{dsfont}
\usepackage[mathscr]{eucal}

\usepackage[normalem]{ulem}
\usepackage{epsfig}
\definecolor{Blue}{rgb}{0.0,0.0,1}
\definecolor{Red}{rgb}{1,0.0,0.0}
\definecolor{Green}{rgb}{0,0.5,0.0}
\usepackage{tikz}
\usetikzlibrary{decorations.pathmorphing}
\usetikzlibrary{arrows}
\usetikzlibrary{intersections,shapes.arrows}
\usetikzlibrary{calc}
\usetikzlibrary{quotes,angles}
\usepackage{nicefrac}
\usepackage{pgfplots}
\usepgfplotslibrary{fillbetween}
\pgfplotsset{compat=1.13,colormap={violetnew}{rgb=(0.293416, 0.0574044, 0.529412) rgb=(0.394818,0.233715,0.671945) rgb =(0.49622,0.410025,0.814477) rgb=(0.588672,0.567494,0.910066) rgb=(0.663226,0.687282,0.911765) rgb=(0.73778,0.807069,0.913465) rgb=(0.807267,0.861883,0.894034) rgb=(0.874222,0.884211,0.864039) rgb=(0.941176, 0.906538, 0.834043)}}
\usepgfplotslibrary{groupplots} 
\usepgfplotslibrary[groupplots] 
\usetikzlibrary{pgfplots.groupplots} 
\usetikzlibrary[pgfplots.groupplots] 
\usepgfplotslibrary{statistics}
\usepackage{pgfplotstable}
\tikzset{jumpdot/.style={mark=*,solid},excl/.append style={jumpdot,fill=white},incl/.append style={jumpdot,fill=black}}
\begin{document}

\title{Probing phase transitions in non-Hermitian systems with Multiple Quantum Coherences}
\author{Diego Paiva Pires}
\affiliation{Departamento de F\'{i}sica, Universidade Federal do Maranh\~{a}o, Campus Universit\'{a}rio do Bacanga, 65080-805, S\~{a}o Lu\'{i}s, Maranh\~{a}o, Brazil}
\author{Tommaso Macr\`{i}}
\affiliation{Departamento de F\'{i}sica Te\'{o}rica e Experimental, Universidade Federal do Rio Grande do Norte, 59072-970 Natal, Rio Grande do Norte, Brazil}
\affiliation{International Institute of Physics, Federal University of Rio Grande do Norte, Campus Universit\'{a}rio, Lagoa Nova, Natal-RN 59078-970, Brazil}

\begin{abstract}
Understanding the interplay between quantum coherence and non-Hermitian features would e\-nable the devising of quantum technologies based on dissipative systems. In turn, quantum coherence can be characterized in terms of the language of multiple quantum coherences (MQCs) originally developed in solid-state nuclear magnetic resonance (NMR), nowadays applied to the detection of quantum chaos, and to the study of criticality in many-body quantum systems. Here we show the usefulness of MQCs for probing equilibrium phase transitions in non-Hermitian systems. To do so, we investigate the connection of quantum coherences and critical points for several paradigmatic non-Hermitian Hamiltonians. For a non-Hermitian two-level system, MQCs witness the parity-symmetry breaking phase transition from the unbroken to the broken phase. Furthermore, for the non-Hermitian transverse field Ising model, MQCs capture the Yang-Lee phase transition in which the ground state energy acquires a nonzero imaginary component. For the disordered Hatano-Nelson (HN) model with periodic boundary conditions, MQCs testify the emergence of mobility edges in the spectrum of this model. In addition, MQCs signal the topological phase transition exhibited by the complex energy spectra of the disorder-free HN model. Finally, we comment on experimentally probing phase transitions in NMR systems realizing non-Hermitian Hamiltonians. Our results have applications to non-Hermitian quantum sensing, quantum thermodynamics, and in the study of the non-Hermitian skin effect.
\end{abstract}

\maketitle


\section{Introduction}
\label{sec:section000000}

Dissipative quantum systems have been widely stu\-died in different contexts~\cite{Plenio1998,Bender2007,Ulrich2012,Rotter,Rotter2015,Bian2021}. Among them, physical systems which can be described by non-hermitian Hamiltonians are particularly important. Indeed, non-Hermitian systems are becoming a central focus of research in optics~\cite{Ruter2010,Zyablovsky_2014,PhysRevA.89.033808,Zyablovsky_2016,Ramy2018,Longhi18,Longhi19,Longhi20,CommunPhys2_37_2019,doi:10.1142/q0178}, photonics~\cite{Longhi2017,Feng2017,NatCommun_9_1308_2018,Klauck2019}, quantum many-body systems~\cite{SMFei_55_1085_2005,Korff_2007,Korff_2008,Castro_Alvaredo_2009,PhysRevE.80.021107,PhysRevLett.125.260601,ProgTheorExpPhys_12_2020,NatCommun_8_15791_2017,Lourenco2018,Xiao2019,PhysRevB.101.121109,PhysRevLett.123.090603,Science_88_6537_2021,PhysRevResearch.2.033022,PhysRevB.102.205118,PhysRevLett.121.203001,Trenkwalder2016,PhysRevLett.123.123601,Yuto_Ashida2020,Nakanishi2021}, quantum metrology~\cite{2020arXiv201101884W,Lau2018,s41467-020-19090-4,PhysRevA.103.042418,Budich2020}, and systems with topological order~\cite{Yuce2018.2,Yuce2019.2,YUCE2020126094,Yao2018,Kunst2019,Song2019.2,Longhi2019.6,Kawabata2019topol,Mittal2020,Kawabata2018,Edvardsson2019,Delplace2021,Bergholtz2021,PhysRevResearch.2.033428,PhysRevB.102.245145,PhysRevX.8.031079,EPJD_74_70_2020,2021_arxiv_2102.03781}, to cite a few. Re\-mar\-kably, some recent experimental realizations shows that non-Hermitian features might stand as a resource for enhancing quantum sen\-sing~\cite{Yu2020,2020arXiv201101884W}.

In addition, theoretical studies have addressed the effects of non-Hermitian driving on quantum co\-he\-ren\-ce~\cite{2103_11496}, also unveiling a mechanism of topological protection of coherence in dissipative quantum systems~\cite{PhysRevA.104.012216}. So far, quantum coherence is a longstanding problem in quantum theory, thus standing as one of its cor\-nersto\-nes~\cite{RevModPhys.89.041003}. It can be fully characterized in terms of coherence orders and multiple quantum co\-he\-rences (MQCs). Overall, both the concepts have been proposed by the community of nuclear magnetic resonance (NMR)~\cite{Munowitz,Keeler} and find applications ran\-ging from entanglement witnessing~\cite{PhysRevA.78.042301} to solid-state spectroscopy~\cite{10.1063_1.449344,MUNOWITZ525,10.1021_ja00284a001,KHITRIN1997217}, and many-body localization~\cite{PhysRevA.84.012320}. Opposite to the technique of quantum state tomography, MQCs require a minimal experimental cost in NMR systems~\cite{Teles}.

Quite recently, non-Hermitian parity-time symmetric Hamiltonians have been successfully implemented in NMR systems~\cite{rsta.2012.0053,PhysRevA.99.062122}. This suggests the investigation of MQCs in nuclear spin systems realizing non-Hermitian Hamiltonians, would shed light on the possible connections between non-Hermitian features and quantum coherence. In a more general scenario, MQC might stand as a {\it bona fide} figure of merit to witness phase transitions in non-Hermitian systems. This is motivated by some recent works showing the MQCs testify quantum phase transitions (QPTs) in Hermitian many-body quantum systems~\cite{PhysRevLett.125.240605}.

In this paper we show the usefulness of the framework of MQCs for probing equilibrium phase transitions in non-Hermitian systems. We address the so-called second moment of multiple quantum intensities (MQIs), thus investigating the link between quantum coherences and critical points for some paradigmatic non-Hermitian Hamiltonians. Overall, our approach is more appealing at both the theoretical and experimental levels. On the one hand, the framework of MQCs is physically meaningful, mostly depending on the ground state coherences of a non-Hermitian system, thus successfully assigning signatures of dissipative equilibrium phase transitions. On the other hand, our results might be implementable with current technology, for example in NMR platforms. In other words, it suggests pro\-bing criticality in non-Hermitian systems by measuring a few elements of the coherence order spectrum that buildup the ground state~\cite{Teles}.

Furthermore, the framework of MQCs is more advantageous when compared with other information-theoretic quantifiers such as quantum fidelity and Loschmidt echoes. We point out that some recent works have proposed to detect equilibrium phase transitions in non-Hermitian quantum systems by using quantum fidelity and Loschmidt echoes~\cite{PhysRevA.94.010102,PhysRevLett.118.015701,PhysRevB.103.155417,PhysRevResearch.3.013015}. Note that such quantifiers mostly depend on the overlap between two states, and its implementation might require preparing copies of the system. Opposite to MQCs, quantum fidelity and Loschmidt echoes require the complete knowledge of the matrix elements of the ground state of the system. In practice, this involves a tomographic state reconstruction task that requires a number of measurements scaling exponentially with the number of particles in a many-body quantum system. In turn, probing criticality in non-Hermitian systems with MQCs would require less information about the system. Indeed, we show that one could infer the dissipative phase transition measuring a few coherence orders of the non-Hermitian ground state respective to some fixed eigenbasis.

For the simplest case of a coupled two-level system with a gain-loss term, we show the second moment of the MQI captures the parity-symmetry-breaking phase transition of this non-Hermitian single-qubit model from the unbroken to the broken phase. In addition, for the non-Hermitian transverse field Ising model, MQIs capture the Yang-Lee phase transition in which the ground state ener\-gy acquires a nonzero imaginary component. For the Hatano-Nelson (HN) model, we show that the second moment of the MQI captures the topological phase transition exhi\-bi\-ted by the complex energy spectra. In detail, this critical behavior occurs in the Hermitian limit of the HN model, i.e., for symmetric hopping amplitudes. In this case, the imaginary part of its eigenenergies become zero, and the phase transition is captured by a sudden change in the winding number~\cite{Kawabata2019topol}, the latter being an integer-valued topological invariant~\cite{PhysRevB.82.235114}. In addition, the MQI testifies the presence of mobility edges in the spectrum of the disordered HN model with periodic boundary conditions (PBCs). Most importantly, measurement of a single coherence order of the ground state for the aforementioned systems can be experimentally performed in current platforms. This can lead to a direct access of non-Hermitian phase transitions for systems whose critical behavior is described by the physical models studied in this paper.

The paper is organized as follows. In Sec.~\ref{sec:section000001} we review useful basic concepts regarding MQCs. In Sec.~\ref{sec:section000002a}, we discuss probing the parity-time-reversal breaking symmetry of non-Hermitian two-level systems. In Sec.~\ref{sec:section000002}, we address the witnessing of Yang-Lee transition with the second moment of the MQI in the next-nearest-neighbor Ising model with complex fields. In Sec.~\ref{sec:section000003}, we show the second moment of the MQI signals the localization of mobility edges in the disordered HN model. In addition, we discuss the pro\-bing of topological phase transitions in the disorder-free HN model. In Sec.~\ref{sec:section000007}, we comment on possible experimental realizations of probing topological phase transitions in non-Hermitian systems in NMR platforms. Finally, in Sec.~\ref{sec:conclusions}, we summarize our conclusions.


\section{Multiple-Quantum Coherences}
\label{sec:section000001}

In this section we briefly review some basic pro\-per\-ties of co\-he\-ren\-ce orders and MQCs. Quantum coherence is a basis-dependent concept, and thus, its characterization requires fixing some preferred basis of states. Let $A$ be an arbitrary observable of a finite-dimensional quantum system, with ${\{|{\psi_{\ell}}\rangle\}_{\ell = 1,\ldots,{2^L}}}$ being its complete set of eigenstates, and ${\{ {\lambda_{\ell}} \}_{\ell = 1,\ldots,{2^L}}}$ the corresponding set of discrete eigenvalues. Hereafter, we will refer to this basis of states as the {\it reference basis}. We furthermore assume that the spacing of the eigenvalue spectrum of $A$ is an integer $m \in \mathbb{Z}$, with ${\lambda_j} - {\lambda_{\ell}} = m$ for all $j,\ell \in \{1,\ldots,{2^L}\}$. In this case, the coherence order decomposition of a quantum state $\rho$ into the reference basis reads
\begin{equation}
\label{eq:00000000000014}
\rho = {\sum_{m}}\, {\rho_m} ~,
\end{equation}
where $\{ {\rho_m} \}_m$ stands as a set of non-Hermitian matrix blocks, with
\begin{equation}
\label{eq:00000000000015}
{\rho_m} = {\sum_{{\lambda_j} - {\lambda_l} = m}}\, {\rho_{jl}}|{\psi_j}\rangle\langle{\psi_l}| ~,
\end{equation}
and ${\rho_{jl}} =  \langle{\psi_j}|\rho|{\psi_l}\rangle$. Then the MQIs are defined as 
\begin{equation}
\label{eq:00000000000017}
{I_m}(\rho) := {\| {\rho_m} \|_2^2} = \text{Tr}({\rho_m^{\dagger}}{\rho_m}) ~.
\end{equation}
We notice that MQI is the Schatten 2-norm, i.e., the Hilbert-Schmidt inner product, of each non-Hermitian block $\rho_m$ in the coherence order decomposition. The se\-cond moment of the MQI of state $\rho$, with respect to the observable $A$, is defined as~\cite{101038nphys4119v01,PhysRevLett.120.040402}
\begin{equation}
\label{eq:00000000000018}
F(\rho,A) = \sqrt{ {\sum_m}\, {m^2}{I_m}(\rho) } ~.
\end{equation}

Importantly, it has been shown that both the MQI spectrum and its second moment can capture signatures of QPTs in many-body quantum systems, also unveiling the role of coherence and entanglement toward the criticality in Hermitian systems~\cite{PhysRevLett.125.240605}. In the following, we will compute the second moment of the MQI for several one-dimensional (1D) non-Hermitian Hamiltonians that can be engineered with ultracold atoms in optical lattices~\cite{ProgTheorExpPhys_12_2020}, dissipative Bose-Einstein condensates~\cite{PhysRevLett.118.045701}, and nuclear spin systems~\cite{PhysRevLett.126.170506}, to cite a few. We will show that, fixing the reference eigenbasis of some observable of the quantum system, the se\-cond moment of the MQI $F(\rho,A)$ for the ground state $\rho$ of the non-Hermitian system stands as a useful quantifier for witnessing non-Hermitian equilibrium phase transition.


\section{Two-level system}
\label{sec:section000002a}

We start by evaluating the MQCs for a paradigmatic two-level system with gain and loss terms. Let $H = (\vec{u} - i\vec{\gamma})\cdot\vec{\sigma}$ be the non-Hermitian two-level system Hamiltonian, where $\vec{u} = \{{u_x},{u_y},{u_z}\}$ and $\vec{\gamma} = \{{\gamma_x},{\gamma_y},{\gamma_z}\}$ are three-dimensional real-valued vectors, while $\vec{\sigma} = ({\sigma_x},{\sigma_y},{\sigma_z})$ is the vector of Pauli matrices. Hereafter we will set ${\| \vec{\alpha} \|^2} := {\sum_l}\, {\alpha_l^2}$ as the Euclidian norm of some vector $\vec{\alpha} = \{{\alpha_x},{\alpha_y},{\alpha_z}\}$. The Hamiltonian exhibits a complex spectrum as $H|{\phi_{\pm}}\rangle = {\kappa_{\pm}}|{\phi_{\pm}}\rangle$, and ${H^{\dagger}}|{\chi_{\pm}}\rangle = {\kappa^*_{\pm}}|{\chi_{\pm}}\rangle$, with ${\kappa_{\pm}} = \pm \|\vec{u} - i\vec{\gamma}\|$, where the set $\{|{\phi_l}\rangle, |{\chi_l}\rangle\}_{l = \pm}$ defines the so-called biorthogonal basis~\cite{Brody_2013}, with the eigenstates given by  
\begin{align}
\label{eq:0000000000001801}
|{\phi_{\pm}}\rangle &= \left( \frac{ {u_z} - i{\gamma_z} \pm \|\vec{u} - i\vec{\gamma} \|}{{u_x} - i{\gamma_x}+ i({u_y} - i{\gamma_y})} \right)|0\rangle + |1\rangle ~, \nonumber\\
|{\chi_{\pm}}\rangle &= \left(\frac{{u_z} + i{\gamma_z} \pm \|\vec{u} + i\vec{\gamma} \|}{{u_x} + i{\gamma_x} + i({u_y} + i{\gamma_y})}\right)|0\rangle + |1\rangle ~, \nonumber\\
\end{align}
where $|0\rangle = {[ 1 \quad 0 ]^{\textsf{T}}}$ and $|1\rangle = {[0 \quad 1]^{\textsf{T}}}$ are the vectors defining the computational basis states in the complex two-dimensional (2D) vector space ${\mathbb{C}^2}$. Importantly, we stress that the biorthogonal basis satisfies the completeness relation ${\sum_{l = \pm}}\, \frac{|{\phi_l}\rangle\langle{\chi_l}|}{\langle{\chi_l}|{\phi_l}\rangle} = \mathbb{I}$, with $\langle{\chi_j}|{\phi_l}\rangle = {\delta_{jl}}\langle{\chi_l}|{\phi_l}\rangle$.

Next, we set the Hermitian o\-pe\-ra\-tor $A = (1/2)(\hat{n}\cdot\vec{\sigma})$ as the ge\-ne\-ra\-tor of unitary evolutions, where $\hat{n} = \{ {n_x},{n_y},{n_z}\}$ is a unit vector with $\| \hat{n} \| = 1$. In this case, the reference basis is composed of the eigenstates $\{ |{\psi_+}\rangle, |{\psi_-}\rangle \}$ of the observable $A$, with
\begin{equation}
\label{eq:0000000000001802}
|{\psi_{\pm}}\rangle = \frac{1}{\sqrt{2}}\left(\pm \sqrt{1 \pm {n_z}} \, |{0}\rangle + \frac{{n_x} + i \, {n_y}}{\sqrt{1 \pm {n_z}}} \, |{1}\rangle\right) ~,
\end{equation}
where we have $A|{\psi_{\pm}}\rangle = {\lambda_{\pm}}|{\psi_{\pm}}\rangle$, with eigen\-va\-lues ${\lambda_{\pm}} = \pm 1/2$. The coherence order spectrum related to the eigenbasis of $A$ is labeled by the set of integers $m = \{ -1,0,+1 \}$. In this case, given the ground state $\rho = |{\phi_-}\rangle\langle{\chi_-}|$ of the non-Hermitian Hamiltonian $H$, its coherence order decomposition reads ${\rho} = {\sum_{m = \{0, \pm 1\}}}\,{\rho_m}$. In turn, the MQI spectrum $\{  {I_{m}}(\rho) \}_{m = 0,\pm 1}$ is obtained from Eq.~\eqref{eq:00000000000017} and reads
 \begin{equation}
\label{eq:0000000000001803}
 {I_{\pm 1}}(\rho) = \frac{[\hat{n}\times(\vec{u} - i\vec{\gamma})]\cdot[\hat{n}\times(\vec{u} + i\vec{\gamma})] \mp 2\,\hat{n}\cdot(\vec{u}\times\vec{\gamma})}{4\, \|\vec{u} - i\vec{\gamma} \| \|\vec{u} + i\vec{\gamma} \|} ~,
 \end{equation}
 and 
  \begin{equation}
\label{eq:0000000000001804}
 {I_{0}}(\rho) = \frac{1}{2}\left( 1 + \frac{[\hat{n}\cdot(\vec{u} - i\vec{\gamma})] \, [\hat{n}\cdot(\vec{u} + i\vec{\gamma})]}{ \|\vec{u} - i\vec{\gamma}\| \|\vec{u} + i\vec{\gamma}\|}  \right) ~.
 \end{equation}
Finally, from Eqs.~\eqref{eq:0000000000001803} and~\eqref{eq:0000000000001804}, the second moment of the MQI spectrum [see Eq.~\eqref{eq:00000000000018}] is written as
\begin{equation}
\label{eq:0000000000001804b}
{F}(\rho,A) = \sqrt{ \frac{[\hat{n}\times(\vec{u} - i\vec{\gamma})]\cdot[\hat{n}\times(\vec{u} + i\vec{\gamma})] }{2\, \| \vec{u} - i\vec{\gamma} \| \| \vec{u} + i\vec{\gamma} \|} } ~.
\end{equation}
Equation~\eqref{eq:0000000000001804} assigns a geometric interpretation to the second moment of the MQI. In fact, we notice that for vectors $\{ \hat{u}, \gamma, \hat{n} \}$ pointing in the same direction along the Bloch sphere, Eqs.~\eqref{eq:0000000000001803} and~\eqref{eq:0000000000001804} lead to the MQI spectrum ${I_{\pm 1}}(\rho) = 0$ and ${I_{0}}(\rho) = 1$, respectively, while Eq.~\eqref{eq:0000000000001804b} implies that ${F}(\rho,A) = 0$. Importantly, this case is equivalent to setting $H$ and $A$ as commuting ope\-ra\-tors, and thus the second moment of the MQI vanishes since the ground state of $H$ is a fully incoherent state with respect to the eigenbasis of the observable $A$.

\begin{figure}[t]
\includegraphics[width=1.05\linewidth]{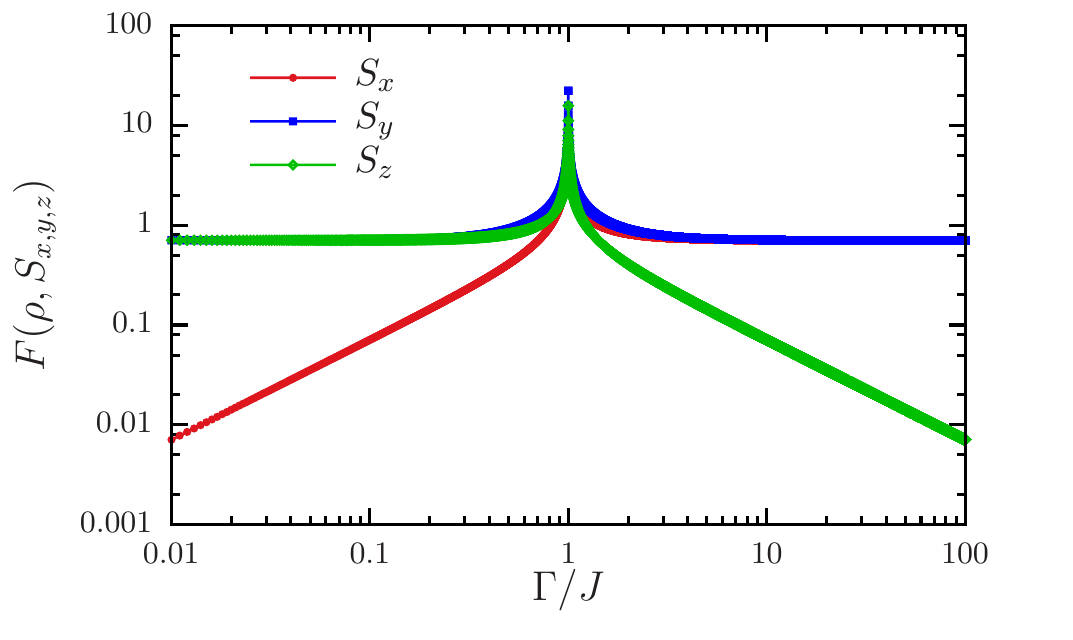}
\caption{(Color online) Second moment of the Multiple-Quantum Intensity spectrum $F(\rho,{S_l})$ for the ground state $\rho = |\phi\rangle\langle{\chi}|$ of the non-Hermitian Hamiltonian $H = J{\sigma_x} + i\Gamma{\sigma_z}$ with respect to the reference eigenbasis of the single-qubit spin operators ${S_{x,y,z}} = (1/2)\, {\sigma_{x,y,z}}$.}
\label{fig000001}
\end{figure}
%
\begin{figure*}[t]
\includegraphics[width=1.00\linewidth]{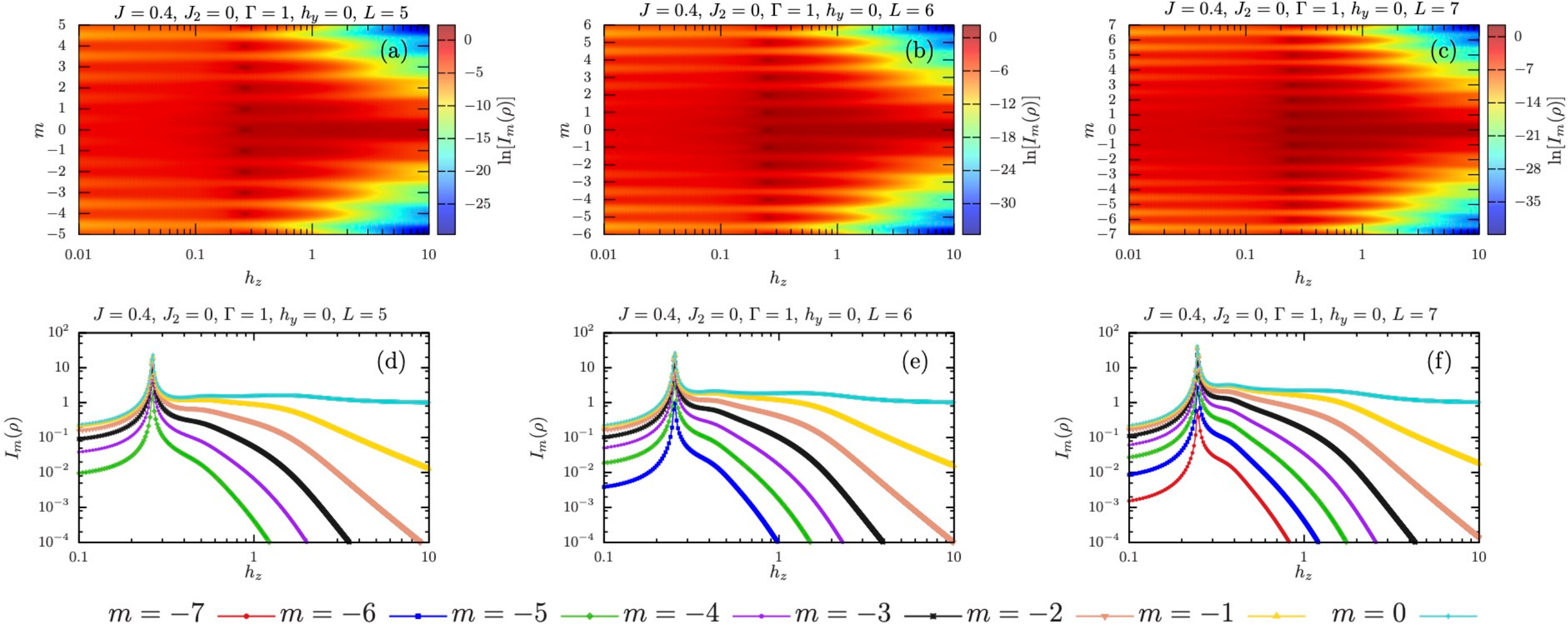}
\caption{Plot of Multiple-Quantum Intensity spectrum $\{{I_m}(\rho)\}_{m = 1,\ldots,L}$ defined in Eq.~\eqref{eq:00000000000017} for the ground state $\rho$ of the non-Hermitian Hamiltonian in Eq.~\eqref{eq:00000000000011} relative to the reference basis of the collective magnetization along the $z$ axis $A = {S_z} = (1/2)\, {\sum_{j = 1}^L}{\sigma_j^z}$. Here, we set $J = 1$, $J_2 = 0$, $\Gamma = 1$, ${h_y} = 0$, and system sizes $L = \{5,6,7\}$ with periodic boundary conditions.}
\label{fig000002}
\end{figure*}
Hereafter, we will specialize our results to the non-Hermitian Hamiltonian $H = J{\sigma_x} + i\Gamma{\sigma_z}$ with parity-time-reversal ($\mathcal{P}\mathcal{T}$) symmetry, with $J$ and $\Gamma$ \-being real parameters~\cite{10.1093ptepptaa181}. This means setting the vectors $\vec{u} = \{J,0,0\}$ and $\vec{\gamma} = \{0,0,\Gamma\}$. Recently, this system has been experimentally rea\-li\-zed in a single trapped ion setup undergoing dissipative perturbations, in which $J$ plays the role of an interlevel coupling strength, while $\Gamma$ is related to a dissipation rate~\cite{PhysRevLett.126.083604}. This system exhibits a phase transition from the unbroken $\mathcal{P}\mathcal{T}$ symmetry-preserving phase ($\Gamma/J < 1$) with real eigenvalues $\pm \sqrt{{J^2} - {\Gamma^2}}$ to the so-called $\mathcal{P}\mathcal{T}$ symmetry-broken phase ($\Gamma/J > 1$), in which the eigenvalues become purely imaginary $\pm i\, \sqrt{{\Gamma^2} - {J^2}}$. At the critical (exceptional) point $\Gamma=J$, the spectrum becomes degenerate with vanishing eigenvalues. To introduce the role of the coherence orders in this non-Hermitian system, we set $\rho = |\phi\rangle\langle{\chi}|$ as the ground state of $H$ and consider the eigenbasis of the collective magnetization operators ${S_{x,y,z}} = (1/2)\, {\sigma_{x,y,z}}$ as the reference basis. It is straightforward to verify from Eq.~\eqref{eq:0000000000001804b} that $F(\rho,{S_x}) = {|\Gamma|} F(\rho,{S_z}) = {|\Gamma|}/{\sqrt{2\, |{J^2} - {\Gamma^2}| }}$, while $F(\rho,{S_y}) = \sqrt{({J^2} + {\Gamma^2})/(2\,|{J^2} - {\Gamma^2}| )}$. 

In Fig.~\ref{fig000001} we plot the second moment of the MQI spectrum $F(\rho,{S_{x,y,z}})$ as a function of the ratio $\Gamma/J$ for the ground state $\rho = |\phi\rangle\langle{\chi}|$ of the aforementioned Hamitonian $H$ relative to the reference eigenbasis of the collective magnetization operators ${S_{x,y,z}} = (1/2)\, {\sigma_{x,y,z}}$. We  notice that $F(\rho,{S_{x,y,z}})$ signals the phase transition between unbroken and broken symmetry sectors. The dissipative phase transition is witnessed by a narrow pick exhibited by $F(\rho,{S_{x,y,z}})$ at the critical point $\Gamma = J$. For the unbroken phase, $\Gamma/J < 1$, note that $F(\rho,{S_{y}})$ and $F(\rho,{S_z})$ saturate around constant values, while $F(\rho,{S_x})$ grows linearly as $\Gamma/J$ increases. However, for the broken phase, $\Gamma/J > 1$, we see that $F(\rho,{S_z})$ decreases, while both $F(\rho,{S_x})$ and $F(\rho,{S_y})$ stay constant as we increase the ratio $\Gamma/J$. Fixing the eigenbasis of $S_{x,y,z}$, the coherences of the ground state of $H$ will take larger values as the system approaches the critical point.


\section{Yang-Lee transition}
\label{sec:section000002}

We now move to a paradigmatic many-body spin system. We consider the $1$D ferromagnetic transverse-field Ising model with next-nearest-neighbor couplings in the presence of imaginary fields~\cite{2021_arxiv_2101.04115}
\begin{equation}
\label{eq:00000000000011}
H = {H_1} + i{H_2} ~,
\end{equation}
with
\begin{equation}
\label{eq:00000000000012}
{H_1} = - {\sum_{j = 1}^L}\, (J{\sigma_j^z}{\sigma_{j + 1}^z} + {J_2}{\sigma_j^z}{\sigma_{j + 2}^z} + {\Gamma}{\sigma_j^x}) ~,
\end{equation}
and
\begin{equation}
\label{eq:00000000000013}
{H_2} = - {\sum_{j = 1}^L}\, ({h_z}{\sigma_j^z} + {h_y}{\sigma_j^y}) ~,
\end{equation}
where ${\sigma^{x,y,z}}$ denotes the Pauli matrices, while the set $\{J,{J_2},\Gamma,{h_y},{h_z}\}$ stand as positive and real-valued parameters. We point out that $[{H_1},{H_2}] \neq 0$, with the Hamiltonian $H$ being non integrable in the absence of imaginary fields for ${J_2} \neq 0$. For $h_y = 0$, the Hamiltonian $H$ belongs to the class of universality of the 2D classical Ising model. This non-Hermitian system undergoes a phase transition that falls into the so-called Yang-Lee universality class, which occurs at ${h_z} \neq 0$ for $\Gamma > J$~\cite{PhysRev.87.404,PhysRev.87.410,PhysRevLett.40.1610,PhysRevLett.54.1354,Gehlen_1991}.

In the following we will compute the MQI spectrum, $\{{I_m}(\rho)\}_m$, and the second moment of the MQI $F(\rho,{S_z})$ for the ground state $\rho$ of Eq.~\eqref{eq:00000000000011} with PBCs. To accomplish this task, we consider the coherence order decomposition of $\rho$ in the eigenbasis of the collective magnetization operator along the $z$ axis, ${S_z} = (1/2){\sum_{j = 1}^L}{\sigma_j^z}$. Unless otherwise stated, we set the parameters $J = 0.4$, $\Gamma = 1$, and ${h_y} = 0$.

In Fig.~\ref{fig000002}, we plot the MQI spectrum for the non-Hermitian Ising model ($J_2 = 0$) as a function of $h_z$, for the system sizes $L = \{ 5,6,7\}$. From Figs.~\ref{fig000002}(a),~\ref{fig000002}(b), and~\ref{fig000002}(c), note that the MQI spectrum is symmetrically distributed around the mode of coherence $m = 0$, which in turn is transla\-tio\-nally invariant regarding unitary rotations generated by the operator $S_z$. We see that, regardless of the size $L$ of the system, the MQI spectrum decreases for larger values of $h_z$, and the amplitudes of ${I_m}(\rho)$ will be mostly dominated by the modes of cohe\-ren\-ce around $m = 0$. Indeed, the higher $h_z$, the less relevant to the MQI spectrum will be to the amplitude of those MQIs ${I_{\pm m}}(\rho)$ related to the coherence orders labeled by integers $m \sim \pm L$.

In Figs.~\ref{fig000002}(d),~\ref{fig000002}(e), and~\ref{fig000002}(f), we see the MQI spectrum $\{{I_m}(\rho)\}_m$ exhibits a narrow peak at some critical point $h_z^c$. Note that, for ${h_z} < {h_z^c}$, the MQI grows as $h_z$ increases, while for ${h_z} > {h_z^c}$, one gets the MQI spectrum decreasing for all integers $m \neq 0$. In turn, the quantity ${I_0}(\rho)$ reaches a stationary value as $h_z$ increases, regardless of the system size $L$, thus being the relevant contribution to the MQI spectrum for $h_z$ away from the critical point. It is worthwhile to note that, the higher the integer $m$, the smaller the amplitude of ${I_m}(\rho)$ as a function of $h_z$. This suggests that, to understand the role of co\-he\-rence orders in this many-body non-Hermitian system, it would suffice to address spin chains including only a few sites, thus evaluating the MQI spectrum around the mode $m = 0$.
\begin{figure}[!t]
\includegraphics[width=1.00\linewidth]{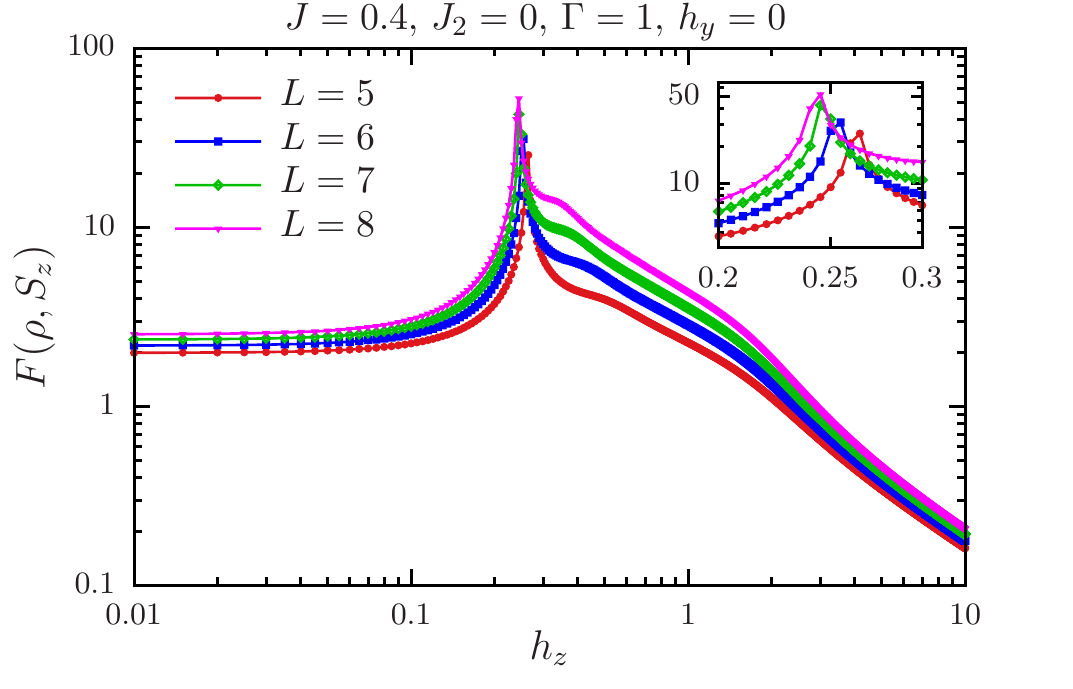}
\caption{Plot of the second moment of the Multiple-Quantum Intensity spectrum $F(\rho,{S_z})$ for the ground state $\rho = |\phi\rangle\langle{\chi}|$ of the non-Hermitian Hamiltonian in Eq.~\eqref{eq:00000000000011} with respect to the reference basis of the collective magnetization operator along the $z$ axis ${S_z} = (1/2){\sum_{j = 1}^L}{\sigma_j^z}$. Here, we set $J = 0.4$, $J_2 = 0$, $\Gamma = 1$, ${h_y} = 0$, and system sizes $L = \{5,6,7,8\}$ with periodic boundary conditions.}
\label{fig000003}
\end{figure}

In Fig.~\ref{fig000003}, we plot the second moment of the MQI $F(\rho,{S_z})$ as a function of $h_z$ for the non-Hermitian integrable Ising model (${J_2} = 0$) with PBCs, for the system sizes $L = \{5,6,7,8\}$. We see that the critical behavior of $F(\rho,{S_z})$ around ${h_z} \approx {h_z^c}$ assigns the phase transition exhibited by the spin model in the paramagnetic regime $\Gamma > J$ for ${h_y} = 0$. The inset shows that the critical point $h_z^c$ slightly decreases as we increase the system size $L$ but starts saturating to a fixed value from $L = 7$ to $8$. Overall, for ${h_z} < {h_z^c}$, note that $F(\rho,{S_z})$ takes some constant value, while it  suddenly grows at the critical point $h_z \approx h_z^c$ and then starts decreasing for ${h_z} > {h_z^c}$. Overall, this means that the larger $h_z$, the less com\-muting will be $H$ and $S_z$. However, for ${h_z} \gg {h_z^c}$, the small values of $F(\rho,{S_z})$ shown in Fig.~\ref{fig000003} indicate that the cohe\-rence orders of the ground state in the eigenbasis of $S_z$ starts decreasing, which means that $H$ and $S_z$ commute.

\begin{figure}[!t]
\includegraphics[width=1.00\linewidth]{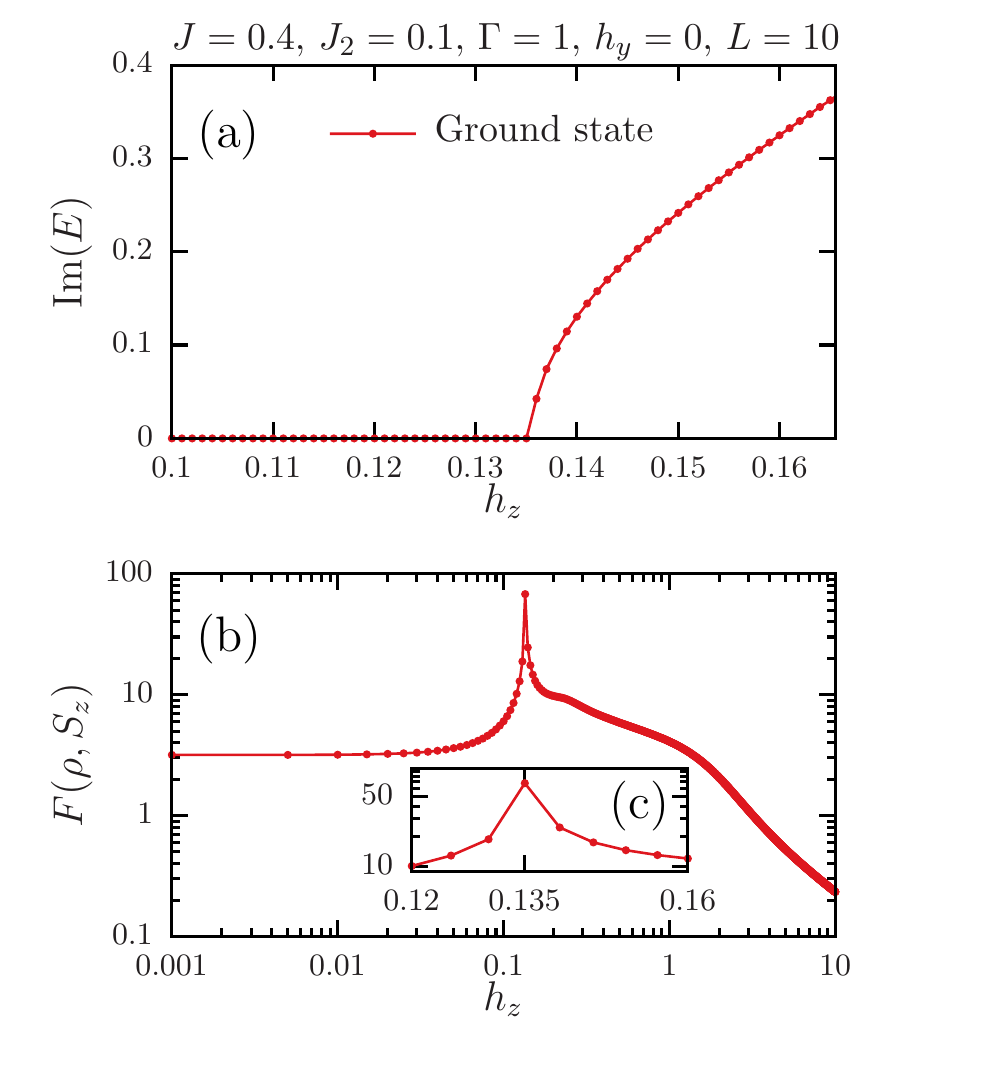}
\caption{Plot of the second moment of the Multiple-Quantum Intensity spectrum $F(\rho,{S_z})$ for the ground state $\rho$ of the non-Hermitian Hamiltonian in Eq.~\eqref{eq:00000000000011} with respect to the reference basis of the collective magnetization operator along the $z$ axis ${S_z} = (1/2){\sum_{j = 1}^L}{\sigma_j^z}$. Here, we set $J = 0.4$, $J_2 = 0.1$, $\Gamma = 1$, ${h_y} = 0$, and system size $L = 10$ with periodic boundary conditions.}
\label{fig000004}
\end{figure}

In Fig.~\ref{fig000004}, we investigate the Yang-Lee edge singu\-la\-ri\-ty for the non-integrable Ising model (${J_2} \neq 0$) with PBCs, setting the spin chain with $L = 10$ sites. In Fig.~\ref{fig000004}(a), we plot the imaginary part $\text{Im}(E)$ of the ground state energy of the system. The Yang-Lee transition is depicted at the critical value ${h_z^c}$ in which the ground state enters a $\mathcal{PT}$ symmetry-broken phase and its energy acquires a nonzero imaginary component, i.e., $\text{Im}(E) > 0$. In this case, we obtain the value ${h_z^c} \approx 0.135$. In Fig.~\ref{fig000004}(b), we plot the second moment of the MQI $F(\rho,{S_z})$ as a function of $h_z$. Importantly, the second moment of the MQI signals the Yang-Lee edge singularity, which in turn is depicted by the narrow peak at the aforementioned cri\-ti\-cal point $h_z^c$. For ${h_z} < {h_z^c}$, the quantity $F(\rho,{S_z})$ saturates around a constant value, while it suddenly increases as the system approaches the critical point, and then starts decreasing for ${h_z} > {h_z^c}$. We clearly see that $F(\rho,{S_z})$ stands as a useful figure of merit for witnessing the phase transition of the non-Hermitian system.

\begin{figure}[!t]
\includegraphics[width=1.0\linewidth]{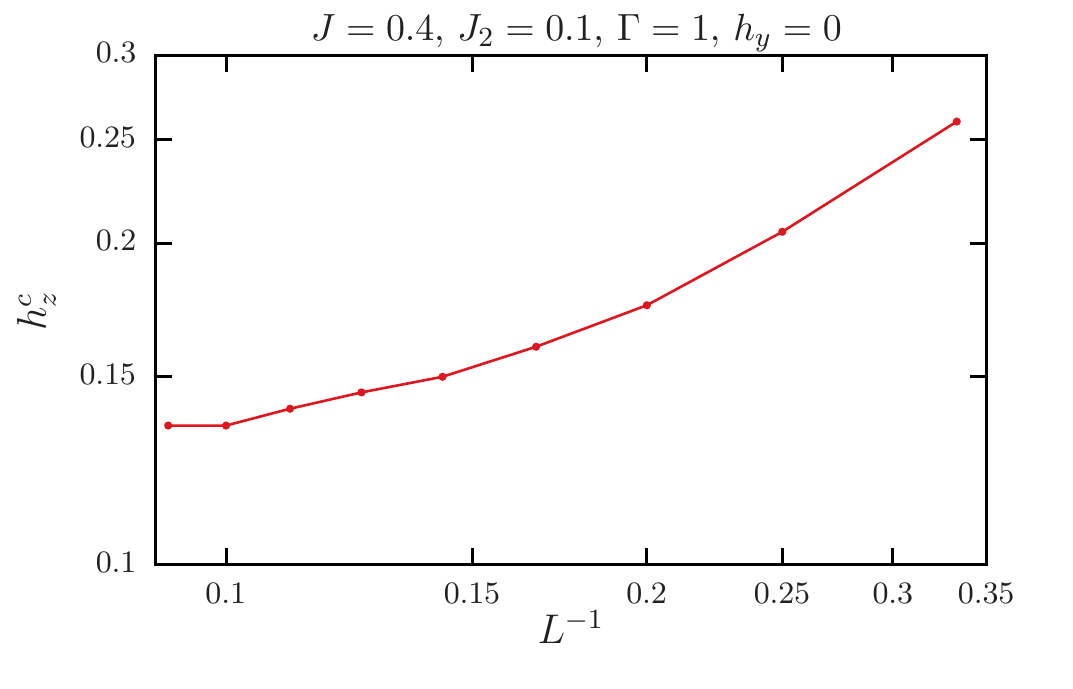}
\caption{Finite-size scaling of the critical points $h_z^c$ of the Multiple-Quantum Intensity spectrum $F(\rho,{S_z})$ as a function of system size $1/L$ for the ground state $\rho = |\phi\rangle\langle{\chi}|$ of the non-Hermitian Hamiltonian in Eq.~\eqref{eq:00000000000011} with respect to the reference basis of the collective magnetization operator along the $z$ axis, ${S_z} = (1/2){\sum_{j = 1}^L}{\sigma_j^z}$. Here, we set $J = 0.4$, $J_2 = 0.1$, $\Gamma = 1$, and ${h_y} = 0$ with periodic boundary conditions.}
\label{fig000005}
\end{figure}

Next, we will discuss the finite-sized scaling of the se\-cond moment of the MQI for the non integrable Ising model with complex fields. In Fig.~\ref{fig000005}, we plot the critical points ${h_z^c}$ of $F(\rho,{S_z})$ as a function of system size $1/L$. We see that the value of the critical point $h_z^c$ mono\-to\-nically decreases as we increase the system size $L$ and starts saturating around a fixed value from $L = 10$ to $11$.


\section{Hatano-Nelson model}
\label{sec:section000003}

Here we consider the so-called disordered Hatano-Nelson model with generalized boundary conditions~\cite{PhysRevLett.77.570,PhysRevB.56.8651,PhysRevB.58.8384,EPJD_74_70_2020}
\begin{align}
\label{eq:hatano000xxx0001}
H &= {\sum_{j = 1}^{N - 1}}\left({J_L}{c_j^{\dagger}}{c_{j + 1}} + {J_R}{c_{j+1}^{\dagger}}{c_j} \right) + {\sum_{j = 1}^{N}}\,{V_j} {c_j^{\dagger}}{c_j} \nonumber\\
&+ {\delta_R}{c_1^{\dagger}}{c_N} + {\delta_L}{c_N^{\dagger}}{c_1}  ~,
\end{align}
where $N$ is the number of lattice sites, ${V_j} \in [-W,W]$ is the on site disorder parameter with $W$ the disorder strength. Here, $J_L$, ${J_R} \in \mathbb{R}$ are imbalanced hopping amplitudes, and ${\delta_L}$, ${\delta_R} \in \mathbb{R}$ determines the generalized boundary conditions. The disorder-free HN model is recovered with $W = 0$, i.e., choosing ${V_j} = 0$ for all $j = \{1,\ldots, N\}$. In addition, note that $\delta_{L,R} \neq 0$ sets the case of generalized PBCs, while for $\delta_{L,R} = 0$ one obtains open boun\-dary conditions (OBCs)~\cite{PhysRevLett.121.026808,Xiong_2018}. In the thermodynamic limit, the complex energy spectrum of $H$ for PBCs display a loop that encircles the origin, with the OBC case comprising a completely real spectrum, regardless of the hopping parameters~\cite{2021_arxiv_2102.03781}. Interestingly, the HN Hamiltonian can be recast as $H = {H_1} + i{H_2}$, with
\begin{align}
\label{eq:hatano000xxx0002}
{H_1} &= \frac{1}{2}(H + {H^{\dagger}}) \nonumber\\
&=  \left(\frac{{J_L} + {J_R}}{2}\right){\sum_{j = 1}^{N - 1}}\left({c_j^{\dagger}}{c_{j + 1}} + {c_{j+1}^{\dagger}}{c_j} \right) \nonumber\\
&+ \left(\frac{{\delta_R} + {\delta_L}}{2}\right)({c_1^{\dagger}}{c_N} + {c_N^{\dagger}}{c_1}) + {\sum_{j = 1}^{N}}\,{V_j} {c_j^{\dagger}}{c_j} ~,
\end{align}
and
\begin{align}
\label{eq:hatano000xxx0003}
{H_2} &= \frac{1}{2i}(H - {H^{\dagger}}) \nonumber\\
&=  \left(\frac{{J_L} - {J_R}}{2i}\right){\sum_{j = 1}^{N - 1}}\left({c_j^{\dagger}}{c_{j + 1}} - {c_{j+1}^{\dagger}}{c_j} \right) \nonumber\\
&+ \left(\frac{{\delta_L} - {\delta_R}}{2i}\right)({c_N^{\dagger}}{c_1} - {c_1^{\dagger}}{c_N}) ~.
\end{align}

In the following, we will discuss the second moment of the MQI $F(\rho,{H_2})$, which in turn signals the coherences of the ground state $\rho = |{\phi}\rangle\langle{\chi}|$ of the HN Hamiltonian with respect to the reference basis of eigenstates of $H_2$. In contrast to spin Hamiltonians, here the eigenvalues of $H_2$ are no longer half-integers, thus implying the index $m$ is an arbitrary real number. Hence, the sum in Eq.~\eqref{eq:00000000000018} is interchanged to run over non-degenerated gaps in the spectrum of $H_2$.

\begin{figure}[!t]
\includegraphics[width=0.9\linewidth]{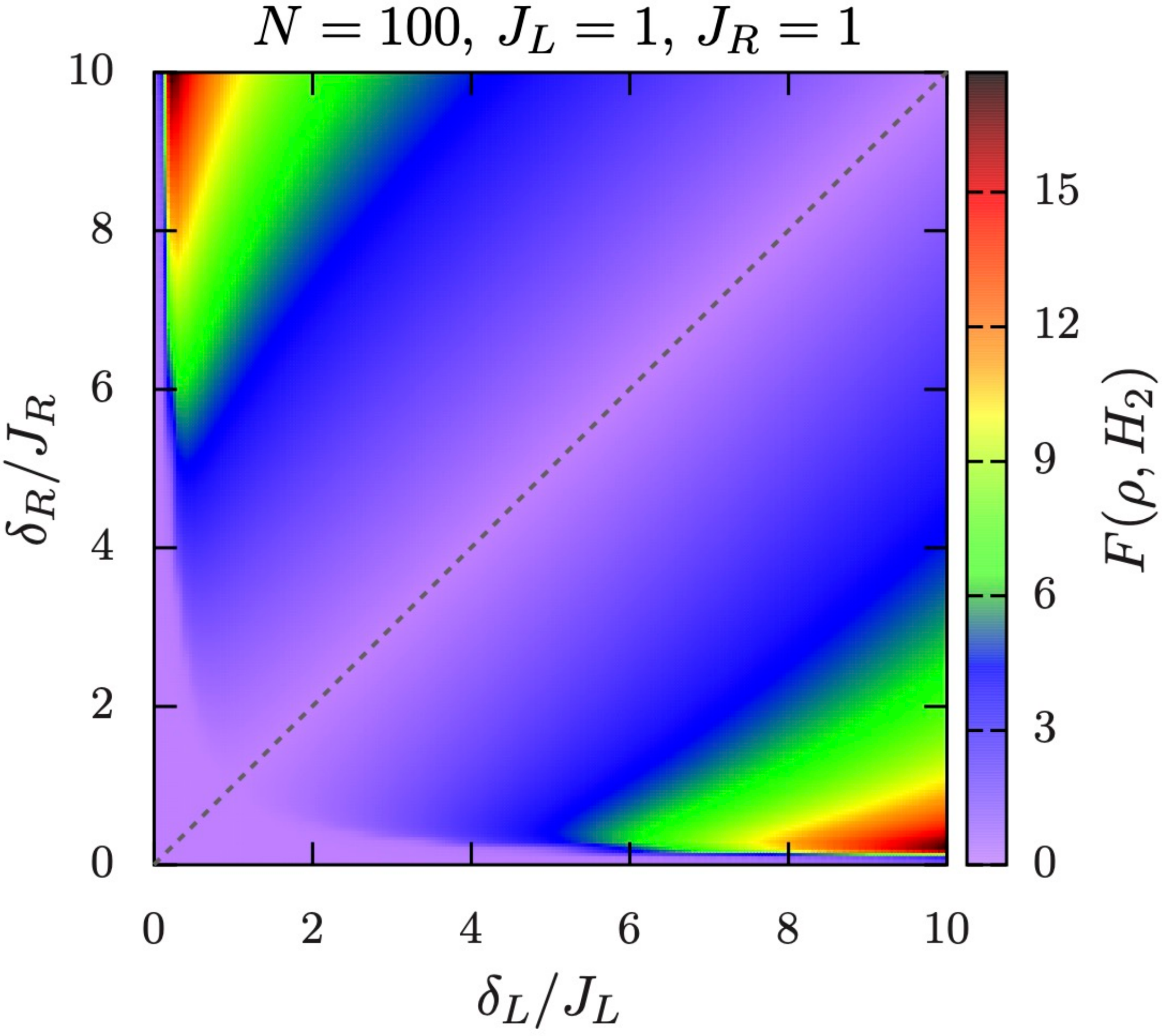}
\caption{Phase diagram of the second moment of the Multiple-Quantum Intensity $F(\rho,{H_2})$ for the ground state $\rho$ of the disorder-free Hatano-Nelson model with periodic boundary conditions [see Eq.~\eqref{eq:hatano000xxx0001}] relative to the reference basis of the Hermitian operator $H_2$ [see Eq.~\eqref{eq:hatano000xxx0003}]. Here, we set the hopping parameters ${J_L} = 1$, ${J_R} = 1$, and system size $N = 100$. Along the gray dashed line (${\delta_L}/{J_L} = {\delta_R}/{J_R}$) $F(\rho,{H_2})$ vanishes.}
\label{fig000006}
\end{figure}


\subsection{Disorder-free Hatano-Nelson model}
\label{sec:section000003AAA}

Here, we address the disorder-free HN model, i.e., we turn off the on site potentials in the Hamiltonian $H$ in Eq.~\eqref{eq:hatano000xxx0001} by setting $W = 0$. For PBCs, one can prove that $H_1$ and $H_2$ are non commuting operators for nonzero parameters $J_L$, $J_R$, $\delta_L$, and $\delta_R$. However, one can verify that $[{H_1},{H_2}] = 0$ for the case ${J_L}/{J_R} = {\delta_L}/{\delta_R}$ (or even ${J_L}/{\delta_L} = {J_R}/{\delta_R}$), which means that $F(\rho,{H_2})$ vanishes since the ground state of $H$ is fully incoherent into the eigenbasis of $H_2$. In the Hermitian limit, i.e., ${J_L} = {J_R}$ and ${\delta_L} = {\delta_R}$, the spectrum of the HN Hamiltonian is no longer complex, and the energies become symmetrically distributed on the real axis, regardless of the system size. In Appendix~\ref{sec:appendix0A} we prove that $F(\rho,{H_2})$ is identically zero for ${\delta_L} = {J_L}$ and ${\delta_R} = {J_R}$. 

In Fig.~\ref{fig000006}, we plot the phase diagram of $F(\rho,{H_2})$ as a function of the ratios ${\delta_L}/{J_L}$ and ${\delta_R}/{J_R}$, setting the hopping parameters ${J_L} = 1$, ${J_R} = 1$, and system size $N = 100$. We see that $F(\rho,{H_2})$ vanishes along the gray dashed line with ${\delta_L}/{J_L} = {\delta_R}/{J_R}$, i.e., both the operators $H_1$ and $H_2$ commute. This critical line depicts a transition between two regions where the ground state is co\-herent in the reference basis. In addition, setting $0 < {\delta_L}/{J_L} \ll 1$ ($0 < {\delta_R}/{J_R} \ll 1$), note the se\-cond moment of the MQI approaches small values for all ${\delta_R}/{J_R} > 0$ (${\delta_L}/{J_L} > 0$), regardless of the system size.
\begin{figure}[t]
\includegraphics[width=1.00\linewidth]{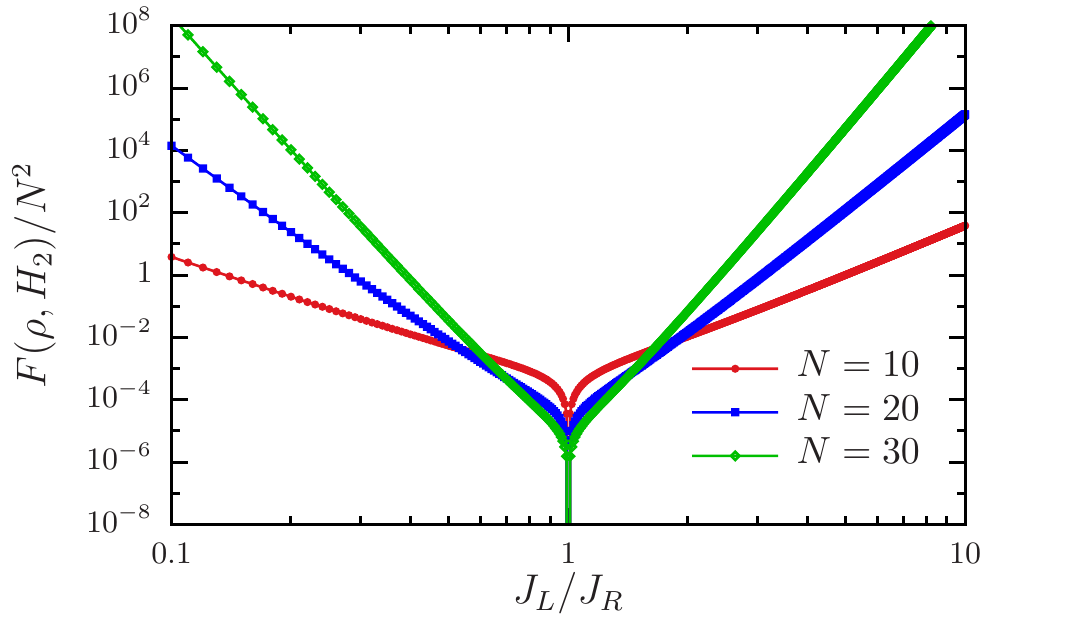}
\caption{Plot of the second moment of Multiple Quantum Intensity $F(\rho,{H_2})$ as a function of the ratio ${J_L}/{J_R}$ for the system sizes $N = \{10,20,30\}$. Here we set the ground state $\rho = |{\psi_N^R}\rangle\langle{\psi_N^L}|$ of the disorder-free Hatano-Nelson model with open boundary conditions (${\delta_{L,R}} = 0$) [see Eq.~\eqref{eq:hatano000xxx0001}] with respect to the reference basis of the Hermitian operator $H_2$ [see Eq.~\eqref{eq:hatano000xxx0003}].}
\label{fig000007}
\end{figure}

Next, for OBCs with ${\delta_{L,R}} = 0$, both the observables $H_1$ and $H_2$ are non commuting ope\-rators for nonzero values of  $J_L$ and $J_R$, except at the exceptional point $J_L = J_R$ in which the spectrum of the HN model undergoes a topological phase transition~\cite{PhysRevX.8.031079}. Hence, we expect that $F(\rho,{H_2})$ signals this critical point by depicting a sudden change between two regions in which $\rho$ is a coherent state regarding the eigenbasis of $H_2$. We will discuss this in more detail. For ${\delta_{L,R}} = 0$, the HN Hamiltonian exhibits the spectral decomposition $H = {\sum_{\ell = 1}^N}\, {E_{\ell}}|{\psi_{\ell}^R}\rangle\langle{\psi_{\ell}^L}|$, where the energy spectrum reads ${E_{\ell}} = 2\, \sqrt{{J_L}/{J_R}} \cos{\phi_{\ell}}$, with ${\phi_{\ell}} = {\ell \pi}/{(N + 1)}$, while the biorthogonal basis is formed by the set of right eigenvectors
\begin{equation}
|{\psi_{\ell}^R}\rangle = \left\{ {c_{\ell,1}}; {c_{\ell,2}}; \ldots ; {c_{\ell,N}} \right\}^{\texttt{T}} ~,
\end{equation}
with
\begin{equation}
{c_{\ell,p}} = \sqrt{\frac{2}{N + 1}} \, {\left(\frac{J_L}{J_R}\right)^{-\frac{p}{2}}} \sin(p{\phi_{\ell}}) ~,
\end{equation}
and the left eigenvectors 
\begin{equation}
|{\psi_{\ell}^L}\rangle = \left\{ {d_{\ell,1}}; {d_{\ell,2}}; \ldots; {d_{\ell,N}} \right\}^{\texttt{T}} ~,
\end{equation}
with
\begin{equation}
{d_{\ell,p}} = \sqrt{\frac{2}{N + 1}} \, {\left(\frac{J_L}{J_R}\right)^{\frac{p}{2}}} \sin(p{\phi_{\ell}}) ~.
\end{equation}

The ground state of the non-Hermitian Hamiltonian $H$ is given by $\rho = |{\psi_N^R}\rangle\langle{\psi_N^L}|$, with energy $E_N = 2\, \sqrt{{J_L}/{J_R}} \cos[N\pi/(N + 1)]$. In the following we will compute the second moment of the MQI $F(\rho,{H_2})$ for the ground state $\rho$ relative to the eigenbasis of the observable $H_2$. From Eq.~\eqref{eq:hatano000xxx0003}, one gets the spectral decomposition ${H_2} = {\sum_{\ell = 1}^N} \, {\widetilde{E}_{\ell}}|{\chi_{\ell}}\rangle\langle{\chi_{\ell}}|$, with the energy spectrum
\begin{equation}
{\widetilde{E}_{\ell}} = \left(\frac{J_L}{J_R} - 1\right)\cos{\phi_{\ell}} ~,
\end{equation}
while the set of eigenstates $\{ |{\chi_{\ell}}\rangle \}_{\ell = 1,\ldots, N}$ is composed of the vectors
\begin{equation}
|{\chi_{\ell}}\rangle = \left\{ {w_{\ell,1}}; {w_{\ell,2}}; \ldots; {w_{\ell,N}} \right\}^{\texttt{T}} ~,
\end{equation}
with 
\begin{equation}
{w_{\ell,p}} = - {i^p}\, \sqrt{\frac{2}{N + 1}} \, \sin(p{\phi_{\ell}}) ~.
\end{equation}
The second moment of the MQI spectrum thus yields
 \begin{equation}
{[F(\rho,{H_2})]^2} =  {\left(\frac{J_L}{J_R} - 1\right)^2} \, {\sum_{j,l = 1}^N}\, {(\cos{\phi_j} - \cos{\phi_l})^2}{|{\rho_{jl}}|^2}  ~,
\end{equation}
with the coefficient ${\rho_{jl}} = \langle{\chi_j}|{\psi_N^R}\rangle\langle{\psi_N^L}|{\chi_l}\rangle$ being the matrix element of the ground state relative to the re\-fe\-ren\-ce basis, which yields
 \begin{equation}
{\rho_{jl}} = \frac{4}{(N + 1)^2} \, {\sum_{p, q = 1}^N}\, {(- 1)^p} \, {i^{p + q}} \, \left(\frac{J_L}{J_R}\right)^{\frac{q - p}{2}} {\xi_{j,l}}(p,q) ~,
\end{equation}
where we define the auxiliary function
\begin{equation}
{\xi_{j,l}}(p,q) =  \sin(p{\phi_N}) \sin(q{\phi_N}) \sin(p{\phi_j}) \sin(q{\phi_l}) ~.
\end{equation}

In Fig.~\ref{fig000007}, we show the second moment of the MQI for the ground state $\rho = |{\psi_N^R}\rangle\langle{\psi_N^L}|$ of the HN model with OBCs, fixing the reference eigenbasis of $H_2$. We set the system sizes $N = \{10,20,30\}$, and plot the quantity $F(\rho,{H_2})/{N^2}$ as a function of the ratio ${J_L}/{J_R}$. We point out that, regardless of the system size, $F(\rho,{H_2})$ vanishes at ${J_L} = {J_R}$, the latter being the exceptional point in which the spectrum of the HN model undergoes a topological phase transition. This critical behavior occurs for the case of symmetric hopping amplitudes in the HN model, and thus, the imaginary part of its eigen\-ener\-gies become zero. It has been shown that this phase transition is witnessed by an abrupt change in the winding number~\cite{Kawabata2019topol}, the latter being an integer-valued topological invariant~\cite{PhysRevB.82.235114}.


\subsection{Disordered Hatano-Nelson model}
\label{sec:section000005}

\begin{figure}[t]
\includegraphics[width=1.0\linewidth]{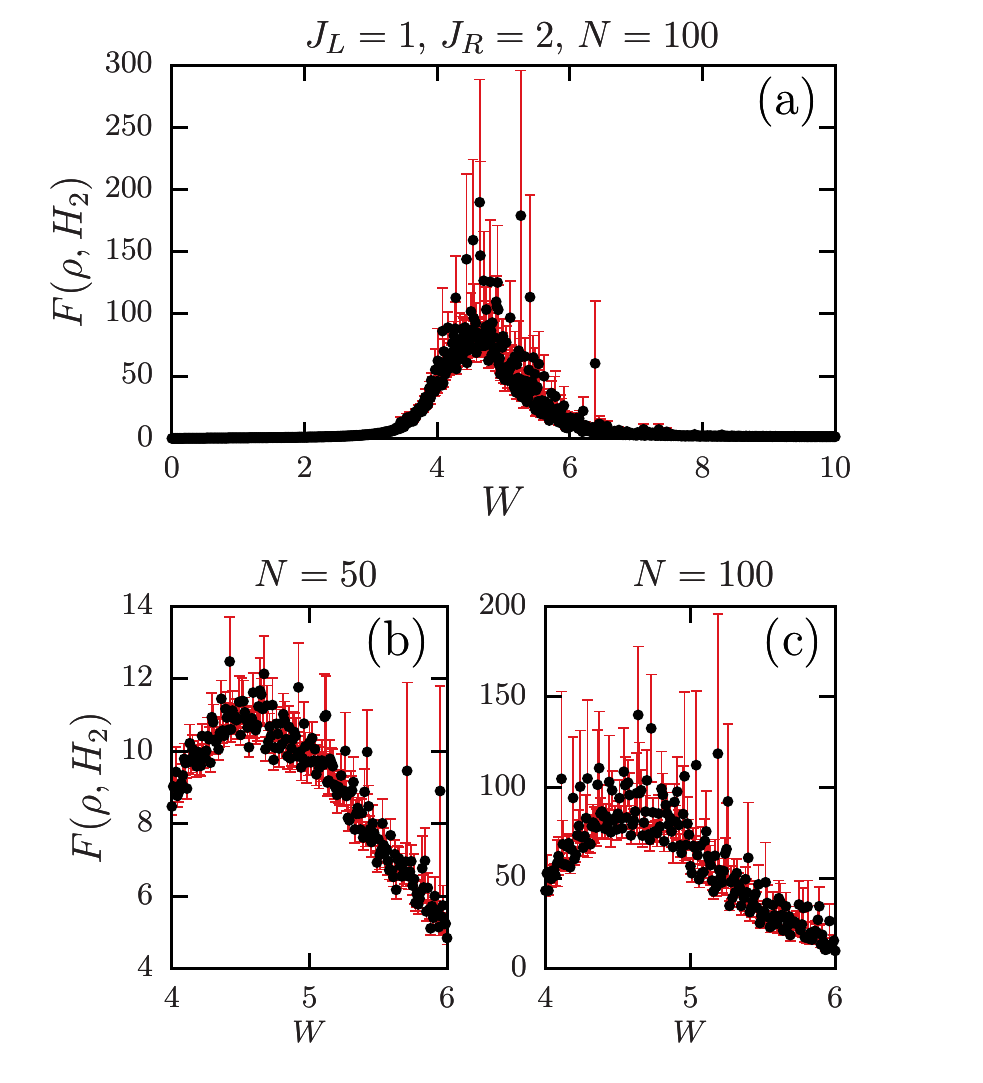}
\caption{Plot of the averaged second moment of the Multiple-Quantum Intensity $F(\rho,{H_2})$ as a function of the disorder strength $W$, for hopping parameters ${J_L} = 1$, ${J_R} = 2$. Here we set the excited state $\rho$ in the middle of the spectrum of the HN Hamiltonian $H$ [see Eq.~\eqref{eq:hatano000xxx0001}] and consider the reference basis of the Hermitian operator $H_2$ [see Eq.~\eqref{eq:hatano000xxx0003}]. The upper panel (a) shows the average of $F(\rho,{H_2})$ over $1100$ realizations for the system size $N =100$, while the lower panels show the ave\-rage of $F(\rho,{H_2})$ over $3000$ realizations, for system sizes (b) $N = 50$, and (c) $N = 100$.}
\label{fig000008}
\end{figure}
Finally, we consider the disordered HN model of Eq.~\eqref{eq:hatano000xxx0001} with disorder strength $W \neq 0$, setting asymmetric hopping amplitudes with ${\delta_L} = {J_L}$ and ${\delta_R} = {J_R}$,~\cite{PhysRevLett.77.570,PhysRevB.56.8651,PhysRevB.58.8384}. In contrast to 1D Hermitian systems that exhibit Anderson localization regardless of the disorder strength~\cite{PhysRevLett.42.673}, the disordered HN model exhibits an Anderson transition~\cite{PhysRev.109.1492}. Importantly, it has been shown that this localization is related to a topological transition in which the complex energy spectrum of the HN model becomes fully real at finite energy. In detail, the mobility edges in the spectrum are mapped onto the origin as $W$ increases, thus increasing the fraction of localized modes in the real axis of the spectrum. Indeed, for ${J_L} = 1$ and ${J_R} = 2$, this transition takes place for disorder strengths $4 \lesssim W \lesssim 6$~\cite{PhysRevX.8.031079}.

In the following, we will investigate localization in the disordered HN model under the viewpoint of coherence orders. We point out that the signa\-ling of localization effects and the buildup of quantum correlations have been addressed under the framework of MQC in many-body quantum systems~\cite{PhysRevLett.104.230403,PhysRevA.84.012320,Alvarez846}. Here, fixing the eigenbasis of $H_2$, we eva\-luate the second moment $F(\rho,{H_2})$ of the MQI, where $\rho = |{\psi^R}\rangle\langle{\psi^L}|$ is an excited state of $H$ [see Eq.~\eqref{eq:hatano000xxx0001}] that is located in the middle of the complex-energy spectra of the disordered HN model.

In Fig.~\ref{fig000008}, we plot the average of $F(\rho,{H_2})$ as a function of the disorder strength $W$ for the hopping parameters ${J_L} = 1$, ${J_R} = 2$. We set the reference basis comprising the eigenstates of the observable $H_2$, thus addressing the coherences of the excited state $\rho$ into this fixed basis. In Fig.~\ref{fig000008}(a), we show the second moment of the MQI a\-ve\-ra\-ged over $1100$ reali\-zations for a system size $N = 100$. We see that $F(\rho,{H_2})$ is almost vanishing for $W \lesssim 3$, while it starts to increase and reaches nonzero values for $W \gtrsim 3$, thus exhibiting a peak around the interval $4 \lesssim W \lesssim 6$. In addition, $F(\rho,{H_2})$ decreases and rapidly approaches zero for $W \gtrsim 6$. This means that, for $4 \lesssim W \lesssim 6$, the excited state becomes coherent in the reference eigenbasis, while $H_1$ and $H_2$ stand as commuting operators for $W \lesssim 3$ and $W \gtrsim 6$. The solid red bars depict the average standard deviation of $F(\rho,{H_2})$ over the disorder realizations, thus showing strong fluctuations around the interval $4 \lesssim W \lesssim 6$, whereas such fluctuations are strongly suppressed for $W \lesssim 3$ and $W \gtrsim 6$. In Figs.~\ref{fig000008}(b) and~\ref{fig000008}(c) we set the system sizes $N = 50$ and $N = 100$, respectively, and plot the quantity $F(\rho,{H_2})$ averaged over $3000$ realizations for $W \in [4,6]$. Overall, the larger the system, the higher the fluctuations on $F(\rho,{H_2})$, which in turn also exhibits higher amplitudes.

Importantly, for typical realizations of the disordered HN model, its complex energy spectra form a loop that encircles the origin for small values of disorder strength~\cite{PhysRevX.8.031079}. In this regard, while the fraction of localized modes increases for $4 \lesssim W \lesssim 6$, Fig.~\ref{fig000008}(a) shows the excited state in the middle of the spectrum exhibits nonzero values of quantum coherences in the eigenbasis of $H_2$. Furthermore, Figs.~\ref{fig000008}(b) and~\ref{fig000008}(c) indicate the line width of the fluctuating peak in $F(\rho,{H_2})$ decreases as we increase the system size. This indicates that $F(\rho,{H_2})$ testifies the interplay of quantum coherences and emergence of the mobility edges in this non-Hermitian disordered system.


\section{Experimental discussion}
\label{sec:section000007}

So far, MQCs have been widely applied for characterizing nuclear spin systems, mostly taking advantage of the ability to control the spins through radiofrequency pulses with NMR spectroscopy~\cite{SUTER1988328,10.1021_ja00284a001,MUNOWITZ525}. Quite recently, non-Hermitian $\mathcal{PT}$-symmetric  Hamiltonians have been implemented in NMR systems~\cite{rsta.2012.0053,PhysRevA.99.062122}. This suggest that NMR platforms might be feasible testbeds for pro\-bing topological phase transitions via the measurement of MQCs. In the following we comment on how to expe\-ri\-mentally probe such phase transitions in non-Hermitian systems by exploiting the framework of MQCs.

Given the ground state ${\rho} = |{\Phi^R_{\text{GS}}}\rangle\langle{\Phi^L_{\text{GS}}}|$ of a non-Hermitian Hamiltonian $H$ spin system, we take $A = {\sum_{l = 1}^N}\, {\lambda_l}|{\psi_l}\rangle\langle{\psi_l}|$ as an observable belonging to the referred physical system, with $\{ |{\psi_l}\rangle \}_{l = 1,\ldots, N}$ being the fixed re\-fe\-rence basis. In this case, one obtains $\rho = {\sum_m}\, {\rho_m}$ as the coherence order decomposition of the ground state in the eigenbasis of $A$, with ${\rho_m} = {\sum_{{\lambda_j} - {\lambda_l} = m}}\, {\rho_{jl}}|{\psi_j}\rangle\langle{\psi_l}|$, and ${\rho_{jl}} =  \langle{\psi_j}|{\rho}|{\psi_l}\rangle$. By hypothesis, the ground state undergoes a unitary evolution $\rho_{\varphi} = {e^{-i\varphi A}} \, {\rho} \, {e^{i\varphi A}} = {\sum_m}\, {e^{- i m \varphi}}{\rho_m}$ that imprints some unknown phase $\varphi$ on it, with $\varphi \in [0,2\pi)$. This process could be experimentally realized in NMR setups by engineering a sequence of microwave pulses that implement collective spin rotations~\cite{SUTER1988328,PhysRevA.80.012328,Cappellaro2014}. Estimating the phase $\varphi$ requires measuring the fidelity $f(\rho,{\rho_{\varphi}}) := \text{Tr}({\rho}{\rho_{\varphi}})$ of the two pure states $\rho$ and $\rho_{\varphi}$. In turn, quantum fidelity relies on the overlap of density matrices~\cite{NJP_22_043001}. In terms of coherence order language, this fidelity is recast as~\cite{101038nphys4119v01,PhysRevLett.120.040402}
\begin{equation}
\label{eq:sec90000001}
f(\rho,{\rho_{\varphi}}) = {\sum_m}\, {I_m}(\rho)\, {e^{- i m \varphi}} ~,
\end{equation}
where ${I_m}(\rho) = \text{Tr}({\rho_m}{\rho_m^{\dagger}})$ is the MQI. Hence, the MQI spectrum can be retrieved from the Fourier transform of this signal, with the $m$-th MQI written as
\begin{equation}
\label{eq:sec90000002}
{I_m}(\rho) = \frac{1}{2\pi}\, {\int_0^{2\pi}}\, d\varphi \, f(\rho,{\rho_{\varphi}})\, {e^{i m \varphi}} ~.
\end{equation}
It is worthwhile to note that both the fidelity and the MQI spectrum have been measured for trapped ions realizing a fully connected Hermitian Ising model~\cite{101038nphys4119v01}.

To probe the information encoded in a given subspace of the quantum system, Eq.~\eqref{eq:sec90000002} means that one can measure the MQI ${I_m}(\rho)$ ruled by the $m$th coherence order sector labeling such quantum subspace. In this case, there is no need to fully reconstruct the ground state of the system via quantum state tomography, the latter exhibiting a complexity that grows exponentially with the number of particles in a many-body quantum system. Re\-mar\-kably, measuring a single MQI typically requires a minimal experimental cost in NMR systems~\cite{Teles}. Hence, this suggests the usefulness of the framework of coherence orders for pro\-bing the ground state of a many-particle quantum system.

To illustrate this idea, we refer to the non-Hermitian transverse field Ising model discussed in Sec.~\ref{sec:section000002}. In detail, given the coherence orders of the ground state related to the collective magnetization operator along the $z$ axis, Figs.~\ref{fig000002}(d),~\ref{fig000002}(e), and~\ref{fig000002}(f) clearly show that the MQI spectrum captures the Yang-Lee transition. In other words, given the $m$-th coherence order, the quantity ${I_m}({\rho})$ unveils an interplay between quantum coherence and the symmetry-breaking phase transition in the many-body system. Hence, one could infer such a phase transition with the measurement of a single MQI, e.g., ${I_0}(\rho)$ that is labeled by the sector $m = 0$. This can be directly applied, for instance, to NMR quantum computing platforms realizing $\mathcal{PT}$-symmetric systems~\cite{rsta.2012.0053,PhysRevA.99.062122}.


\section{Conclusions}
\label{sec:conclusions}

In conclusion, we have shown the usefulness of the coherence order framework for probing phase transitions in non-Hermitian systems. Focusing on the second moment of MQIs, we ve\-ri\-fied the interplay of quantum coherences and critical points for some prototypical non-Hermitian Hamiltonians.

For the ground state of non-Hermitian two-level systems, fixing a given reference basis, we have shown that the second moment of the MQI displays a critical behavior at the same critical points as the spectrum of non-Hermitian system (see Fig.~\ref{fig000001}). In other words, measuring the coherences of the ground state respective to the fixed eigenbasis, the second moment of the MQI witnesses the parity-symmetry breaking phase transition of the single-qubit non-Hermitian model.

Next, for the non-Hermitian transverse field Ising model with PBCs, the second moment of the MQI for its ground state exhibits a critical behavior as a function of the magnetic field $h_z$, thus displaying the Yang-Lee phase transition (see Fig.~\ref{fig000004}). We see that this critical behavior persists for both the (Hermitian) integrable and non integrable cases (see Fig.~\ref{fig000003}). For the latter, the finite-sized scaling of the second moment of the MQI shows that the critical point monotonically decreases as a function of the system size $L^{-1}$ and starts saturating around a fixed value for $L \gtrsim 10$ (see Fig.~\ref{fig000005}). Importantly, we see the MQI spectrum signals this critical point, and thus one could probe the non-Hermitian phase transition by measuring a single coherence order of the ground state (see Fig.~\ref{fig000002}).

For the HN model with OBCs, we have shown that the second moment of the MQI captures the topological phase transition exhibited by the complex energy spectra (see Fig.~\ref{fig000007}). Indeed, the MQI vanishes at the exceptional point displaying the Hermitian limit of the HN model. For PBCs, the second moment of the MQI depicts two nonzero regions where the ground state is co\-he\-rent in the fixed reference basis. 

We have verified the second moment of the MQI unveils a signature of the emergence of mobility edges in the spectrum of the disordered HN model with PBCs. We have shown that, for some excited state in the middle of the complex energy spectra, the ave\-raged second moment of the MQI mostly vanishes, except for a peak that appears for a given range of the disorder strength (see Fig.~\ref{fig000008}(a)). This peak has a strongly fluctuating amplitude, and its width decreases as we increase the system size (see Figs.~\ref{fig000008}(b) and~\ref{fig000008}(c)). Importantly, this peak occurs around some values of disorder strength for which it is known the whole spectrum becomes localized. We expect that, increasing both the system size and the number of averaging realizations, the peak will become more pronounced, while the disorder strength approaches the critical value for localization transition of mobility edges in the disordered HN model.

Finally, we discussed an experimentally relevant scheme to probe equilibrium phase transitions in non-Hermitian systems by exploiting the framework of MQCs. Our results suggest that one could probe criticality in non-Hermitian systems by measuring a few elements of the MQI spectrum, the latter ruling the coherence orders that buildup the density matrix. The results in this paper could find applications in the subject of non-Hermitian quantum thermodynamics~\cite{deffner_scirep_23408_2016} and in the study of enhancing quantum sensing with non-Hermitian systems~\cite{s41467-020-19090-4,PhysRevA.103.042418}. 


\begin{acknowledgments}
D. P. P. and T. M. acknowledge the financial support from the Brazilian ministries MEC and MCTIC and fun\-ding agencies CAPES and CNPq. T. M. acknowledges CNPq for support through Bolsa de produtividade em Pesquisa n.311079/2015-6. T. M. was supported by the Serrapilheira Institute (Grant No. Serra-1812-27802), CAPES-NUFFIC Project No. 88887.156521/2017-00.
\end{acknowledgments}

\setcounter{equation}{0}
\setcounter{figure}{0}
\setcounter{table}{0}
\setcounter{section}{0}
\numberwithin{equation}{section}
\makeatletter
\renewcommand{\thesection}{\Alph{section}} 
\renewcommand{\thesubsection}{\thesection.\arabic{subsection}}
\renewcommand{\theequation}{\Alph{section}\arabic{equation}}
\renewcommand{\thefigure}{\arabic{figure}}
\renewcommand{\bibnumfmt}[1]{[#1]}
\renewcommand{\citenumfont}[1]{#1}


\section*{Appendix}


\section{HN model with periodic boundary conditions}
\label{sec:appendix0A}

Let us consider the HN model with PBCs, also setting ${\delta_L} = {J_L}$ and ${\delta_R} = {J_R}$, whose Hamiltonian reads as $H = {H_1} + i{H_2}$, where
\begin{equation}
\label{eq:APPENDIX0000001}
{H_1} =  \left(\frac{{J_L} + {J_R}}{2}\right){\sum_{j = 1}^{N}}\left({c_j^{\dagger}}{c_{j + 1}} + {c_{j+1}^{\dagger}}{c_j} \right) ~,
\end{equation}
and
\begin{equation}
\label{eq:APPENDIX0000002}
{H_2} =  \left(\frac{{J_L} - {J_R}}{2i}\right){\sum_{j = 1}^{N}}\left({c_j^{\dagger}}{c_{j + 1}} - {c_{j+1}^{\dagger}}{c_j} \right) ~,
\end{equation}
with the constraint $c_{N + 1} = c_1$. Here we will show the se\-cond moment of the MQI, i.e., $F(\rho,{H_2})$, identically vanishes when considering the ground state $\rho$ of the HN model, fixing the reference eigenbasis of $H_2$. To do so, note the non-Hermitian HN Hamiltonian exhibits the complex spectrum given by~\cite{2021_arxiv_2102.03781}
\begin{equation}
\label{eq:APPENDIX0000003}
{E_n} = \left(\frac{J_L}{J_R} + 1\right)\cos{\theta_n} + i \left(\frac{J_L}{J_R} - 1\right)\sin{\theta_n} ~,
\end{equation}
with $\theta_n = {2\pi n}/{N}$, and also the set of eigenvectors $\{ |{\psi_n}\rangle \}_{n = 1,\ldots,N}$, with
\begin{equation}
\label{eq:APPENDIX0000004}
|{\psi_n}\rangle = \frac{1}{\sqrt{N}}\left\{ {e^{i{\theta_n}}} , {e^{2i{\theta_n}}} , \ldots , {e^{Ni{\theta_n}}} \right\}^{\texttt{T}} ~.
\end{equation}
For $N$ even, the ground state of the Hamiltonian $H$ is labeled as $n = N/2$. However, for $N$ odd, the ground state exhibits a two-fold degeneracy and is obtained for $n = (N \pm 1)/2$. We point out that $H_2 = {\sum_n}\, {\mathcal{E}_n} |{\phi_n}\rangle \langle{\phi_n}|$ is the spectral decomposition of $H_2$, with the real energies
\begin{equation}
\label{eq:APPENDIX0000005}
{\mathcal{E}_n} = \left(\frac{J_L}{J_R} - 1\right)\sin{\theta_n} ~,
\end{equation}
and the eigenstates
\begin{equation}
\label{eq:APPENDIX0000006}
|{\phi_n}\rangle = \frac{ {(-1)^n}}{\sqrt{N}}\left\{ { e^{i{\theta_n}}} , {e^{2i{\theta_n}}} , \ldots ,  {e^{Ni{\theta_n}}} \right\}^{\texttt{T}} ~.
\end{equation}

For $N$ even, it follows that $\rho = |{\psi_{{N}/{2}}}\rangle\langle{\psi_{{N}/{2}}}|$ stands as the ground state of the HN Hamiltonian. In this case, the second moment of the MQI spectrum thus yields
 \begin{equation}
 \label{eq:APPENDIX0000007}
{[F(\rho,{H_2})]^2} =  \frac{1}{N^2} {\left(\frac{J_L}{J_R} - 1\right)^2} \, {\sum_{j,l = 1}^N}\, {(\sin{\theta_j} - \sin{\theta_l})^2}{|{\rho_{jl}}|^2} ~,
\end{equation}
where ${\rho_{jl}} := \langle{\phi_j}|{\psi_{{N}/{2}}}\rangle\langle{\psi_{{N}/{2}}}|{\phi_l}\rangle$ is the matrix element of the ground state with respect to the eigenbasis of $H_2$, and can be recast as
\begin{align}
\label{eq:APPENDIX0000008}
{\rho_{jl}} &= {(-1)^{j + l}} \left( \frac{1}{N} \, {\sum_{p = 1}^N}\, {e^{-\frac{2\pi i p}{N}\left(j - \frac{N}{2}\right)}} \right)\left( \frac{1}{N}\,  {\sum_{q = 1}^N}\, {e^{-\frac{2\pi i q}{N}\left(l - \frac{N}{2}\right)}} \right) \nonumber\\
&= {(-1)^{j + l}} \, {\delta_{j,N/2}}\, {\delta_{l,N/2}} ~,
\end{align}
where we have recognized the product of the Kronecker delta. Hence, plugging Eq.~\eqref{eq:APPENDIX0000008} into Eq.~\eqref{eq:APPENDIX0000007}, it is straightforward to conclude the second moment of the MQI vanishes, i.e., ${F(\rho,{H_2})} =  0$ for $N$ even. Next, for $N$ odd, the ground state of $H$ is two-fold degenerated as $\rho = |{\psi_{\frac{N \pm 1}{2}}}\rangle\langle{\psi_{\frac{N \pm 1}{2}}}|$. The second moment of the MQI spectrum thus yields
 \begin{equation}
 \label{eq:APPENDIX0000009}
{[{F^{\pm}}(\rho,{H_2})]^2} =  {\left(\frac{J_L}{J_R} - 1\right)^2} \, {\sum_{j,l = 1}^N}\, {(\sin{\theta_j} - \sin{\theta_l})^2}{|{\rho^{\pm}_{jl}}|^2}  ~,
\end{equation}
where here ${\rho_{jl}^{\pm}} := \langle{\phi_j}|{\psi_{\frac{N \pm 1}{2}}}\rangle\langle{\psi_{\frac{N \pm 1}{2}}}|{\phi_l}\rangle$, which is written as
 \begin{align}
 \label{eq:APPENDIX0000010}
{\rho^{\pm}_{jl}} &=  {(-1)^{j + l}} \left( \frac{1}{N} \, {\sum_{p = 1}^N}\, {e^{\frac{2\pi i p}{N}\left(\frac{N \pm 1}{2} - j\right)}} \right)\left( \frac{1}{N}\,  {\sum_{q = 1}^N}\, {e^{\frac{2\pi i q}{N}\left(l - \frac{N \pm 1}{2}\right)}} \right) \nonumber\\
&= {(-1)^{j + l}} \, {\delta_{j, \frac{N \pm 1}{2}}}\, {\delta_{l, \frac{N \pm 1}{2}}} ~.
\end{align}
Finally, substituting Eq.~\eqref{eq:APPENDIX0000010} into Eq.~\eqref{eq:APPENDIX0000009}, we see the second moment of the MQI vanishes, i.e., ${F(\rho,{H_2})} =  0$ for $N$ odd. As a final comment, we point out that ${F(\rho,{H_2})}$ is expected to be zero since $H_1$ and $H_2$ are commuting o\-pe\-ra\-tors, and thus the ground state of $H$ is an incoherent state with respect to the eigenbasis of $H_2$.

%

%

%
%
%
%
\end{document}